\documentstyle[preprint,prd,aps]{revtex}
\begin{document}
\draft
\preprint{RU9609, USC(NT)-96-2}
\title{Amplitude analysis of the $\overline{N}N \rightarrow 
K^-K^+$ reaction } 
\author{W.M. Kloet}
\address{Department of Physics and Astronomy, Rutgers University, \\
Piscataway, New Jersey 08855, USA}
\author{F. Myhrer}
\address{Department of Physics and Astronomy, University of South Carolina, \\
Columbia, South Carolina 29208, USA \\}
\date{\today}
\maketitle
\begin{abstract}
A simple partial wave amplitude analysis of 
$\overline{p}p \rightarrow K^-K^+$ 
has been performed for data in the range $p_{\rm lab}$ = 360 -- 
1000 MeV/$c$.
In this low momentum interval only partial wave amplitudes 
with $J$ equal to 0, 1 and 2 are needed to obtain a good fit to 
the experimental data. 
This maximal $J$ = 2 value is smaller than what 
is required for the data of the reaction
$\overline{p}p \rightarrow \pi^-\pi^+$ 
in the same momentum interval. \\
\end{abstract}


\narrowtext

The reaction $\overline{N}N \rightarrow K^-K^+$ differs from the 
reaction $\overline{N}N \rightarrow \pi^-\pi^+$
in that at least two pairs of valence quarks and antiquarks 
must be annihilated and 
a strange quark  antiquark pair must be created. 
We expect the $\overline{p}p \rightarrow K^-K^+$
reaction to take place in a smaller interaction volume than the 
$\overline{N}N \rightarrow \pi^-\pi^+$ reaction. 
A reason for this expectation is illustrated by 
the analogy with the atomic annihilation processes of 
$\mu^+ e^-$ and  $\mu^- e^+$ into three possible final states 
which are respectively 
$\mu^+ \mu^- + e^+ e^-$ (rearrangement),
$\mu^+ \mu^- + \gamma$ (one lepton pair annihilation) and 
$\gamma + \gamma$ (two lepton pair annihilation). 
In this atomic case there is a sharp decrease in interaction volume 
for the three respective mechanisms of annihilation, 
as mentioned for example in Ref. \cite{richard}. 
Since such arguments are based on QED and 
not on strong interactions this analogy 
should only be taken as an indication of what might be occurring
when comparing the two annihilation channels of $\overline{N}N$ 
into $K^- K^+$ and $\pi^- \pi^+$, 
which is the topic of the present paper. 
If this analogy has merit, we expect that, in 
terms of a partial wave analysis of 
$\overline{p}p \rightarrow K^-K^+$, fewer partial
waves will be active in the $K^-K^+$ annihilation channel as 
compared to the $\pi^-\pi^+$ channel. 
This feature is not readily apparent in the 
data for $\overline{p}p \rightarrow \pi^- \pi^+$
and $\overline{p}p \rightarrow K^- K^+$ 
from the CERN Low Energy Antiproton Ring (LEAR)~\cite{LEAR}.  
At the lowest energies the measured 
$ d \sigma /d \Omega$ and analyzing power $A_{on}$ 
show a rather rich angular dependence 
in both reactions. This dependence changes rapidly with 
increasing energy. 
A detailed analysis of these very good LEAR data may shed some 
light on the microscopic  annihilation and subsequent 
hadronization processes involved and guide us in obtaining a 
model understanding of these elementary annihilation reactions. 

In this paper we report on a simple 
amplitude analysis of the $\overline{p}p \rightarrow K^- K^+$ 
data in the restricted momentum range, $p_{\rm lab}$ = 360 -- 
988 MeV/$c$. 
The method of analysis and the assumptions are  
the same as were used in the reaction $\overline{N}N \rightarrow 
\pi^-\pi^+$~\cite{Klo}. 
Because we know of no reliable model for 
any of these two annihilation processes to guide or to restrict 
this analysis, 
we reduce the theoretical input of this analysis to a minimum. 
We assume only that very few partial waves
contribute to this annihilation reaction. 
This means that only partial waves with $J$ smaller or equal 
to $J_{max}$ ($J_{max}$ is one of the parameters in this analysis) 
contribute to the observables.
In a previous paper we analyzed  the 
$\overline{N}N \rightarrow \pi^-\pi^+$
reaction~\cite{Klo} based on a $\chi^2$ fit and we found 
that all the amplitudes with $J \leq$ $J_{max}$ = 3  
were necessary and sufficient to fit the data below 1 GeV/$c$. 

Similar to the reaction $\overline{N}N \rightarrow \pi^-\pi^+$
there are also two independent helicity amplitudes
$\rm f_{++}$ and $\rm f_{+-}$ for the annihilation reaction 
$\overline{p}p \rightarrow K^- K^+$.
The two measured observables are the differential cross section
\begin{equation}
d\sigma/d\Omega = (|f_{++}|^2 + |f_{+-}|^2)/2,
\end{equation}
and the analyzing power $A_{on}$, defined by

\begin{equation}
{\sl A}_{\sl on} {\sl d}\sigma/{\sl d}\Omega = Im (f_{++} \: f_{+-}^* ).
\end{equation}
The convention is used where {\bf \^{n}} is the spin direction 
normal to the scattering plane. The unit-vector {\bf \^{n}} is along
$\vec{p}$ x $\vec{q}$, where $\vec{p}$ is the antiproton 
center-of-mass (c.m.) momentum and $\vec{q}$ is the c.m. momentum of 
$K^-$.

Additional spin observables  are the spin-correlations 
$A_{\sl ss}$ and $A_{\sl ls}$, 
which are expressed by the following helicity amplitude combinations:
\begin{equation}
{\sl A}_{\sl ss} \: {\sl d}\sigma/{\sl d}\Omega = (|f_{++}|^2 - |f_{+-}|^2)/2 ,                 
\end{equation}
and 

\begin{equation}
{\sl A}_{\sl ls} \: {\sl d}\sigma/{\sl d}\Omega = {Re} (f_{++} \: f_{+-}^* ) .
\end{equation}
However,  there are as yet no data on spin-correlations for this 
reaction.

The two helicity amplitudes are expanded in
$J \ne L$ spin-triplet partial wave amplitudes
\begin{equation}
 f_{++} = \frac{1}{p} \sum_J \sqrt{J+\frac{1}{2}} \: ( 
\sqrt{J}\:
f_{J-1}^J    - \sqrt{J+1} \: f_{J+1}^J ) P_J(cos \theta)
\end{equation}
and

\begin{equation}
 f_{+-} = \frac{1}{p} \sum_J \sqrt{J+\frac{1}{2}} ( 
\frac{1}{\sqrt{J}} \:
 f_{J-1}^J  + \frac{1}{\sqrt{J+1}} \: f_{J+1}^J )  
P_J^{\prime}(cos \theta) sin \theta , 
\end{equation}
where $p$ is the $\overline{p} p$ center of mass momentum and
where $P_J^{\prime}$
denotes the first derivative of the Legendre polynomial $P_J$.

In the data analysis of $\overline{p}p \rightarrow
K^-K^+$, we parameterize the partial wave amplitudes at each 
energy as

\begin{equation}
 f_{\sl L}^{\sl J} = R_{\sl LJ} \: e^{\sl i \: \delta_{\sl 
LJ}}.
\end{equation}
where $R_{\sl LJ}$ and $\delta_{\sl LJ}$ are free parameters.
At each energy we choose the  $J_{max}$ to be included in our
$\chi ^2$ search and find the best fit to both
$\rm {\sl d}\sigma/{\sl d}\Omega$ and $A_{\sl on}$ 
by minimizing the $\chi ^2$ sum.
In our fits we  choose
$\delta_{10}$ = 0  for the $^3P_0$ partial wave
whereas $R_{10}$ is a free parameter.
For all other $LJ$ values both phase and amplitude in Eq. (7)
are allowed to vary to obtain the best fit.
In our analysis we did not try to correlate the 
amplitudes at the different energies by a smoothness procedure.
We did however use the set of amplitude values, 
determined at one energy, 
as start values in the $\chi^2$ search at the neighboring energy.
Due to the incompleteness of the set of measured observables,
we do not find a unique solution, i.e., a unique
set of partial wave amplitudes in our $\chi^2$ search.
However, the minimal values of $\chi^2$ 
found in the various possible fits are the same. 
As discussed in our  analysis of 
$\overline{N}N \rightarrow \pi^-\pi^+$   \cite{Klo}, if      
we had available data on other spin observables
it would be possible to restrict the choice among the various
amplitude-sets with equally good $\chi^2$.

The data for all measured energies
starting from $p_{\rm lab}$ = 360 MeV/$c$ up to 1  GeV/{\sl c}
can be fitted with partial wave amplitudes with total angular 
momentum $J$ $\leq$ 2. We have also fitted the data with a maximal 
$J$ = 3 using the same procedure.
It appears that for $\overline{p}$ momenta, $p_{\rm lab}$, above 
886 MeV/$c$ the total $\chi^2$ improves when we include
the $J$ = 3 partial wave amplitudes, but below 886 MeV/$c$
the improvement is marginal.
As examples we show three fits in Figs.~1-3 for respectively 
$p_{\rm lab}$ = 360, 585, and 988 MeV/$c$. 
By including the $J$ = 3 amplitude the
$\chi^2$ per degree of freedom hardly improves 
except for $d\sigma/d\Omega$ at 988 MeV/$c$ as seen in Fig.~3.
It is remarkable that so few partial waves with $J_{\sl  max}$ = 2 are 
sufficient in order to get a satisfactory $\chi^2$  fit to the data 
in such a large energy range.
On the other hand, we note that the $J$ = 2 partial wave amplitude is  
essential already at the three lowest measured energies
due to the presence of two minima in the angular behaviour  
of both the differential cross section 
$d\sigma/d\Omega$ and of the asymmetry $A_{\sl on}$.

In Table~I we show the $\chi^2$ per degree of freedom for
one set of partial wave amplitude parameters with $J_{\sl  max}$ 
= 2 as well as the case
where $J_{\sl  max}$ =  1, 3 or 4. Listed are the ten momenta
between $p_{\rm lab}$ = 360 and 988 MeV/$c$, where there are 
available LEAR data. From Table I one notes that $\chi^2$ 
for $J_{\sl max}$ = 2 is much lower than the corresponding $\chi^2$ 
for $J_{\sl max}$ = 1. 
On the other hand an increase of the value of  $J_{\sl max}$ to 3 
or to 4 does not improve the $\chi^2$ significantly except at 
$p_{\rm lab}$ = 988 MeV/$c$. 
For comparison we give in Table~II a similar list of 
$\chi^2$ for the process 
$\overline{p}p \rightarrow \pi^-\pi^+$~\cite{Klo}~.
In that case the preferred maximum angular momentum 
is clearly $J_{\sl max}$ = 3 at all momenta. 
The values of $\chi^2$ for 
$\overline{p}p \rightarrow K^-K^+$ 
are less smooth than for 
$\overline{p}p \rightarrow \pi^-\pi^+$, 
because for the $K^-K^+$ final state 
there are only half as many data points and in addition the 
error bars in the data are larger than for annihilation 
into $\pi^-\pi^+$. 
However, given our simple amplitude assumptions and the large 
experimental errors for $A_{\sl on}$, 
we did not make a serious error analysis of our $\chi^2$ fit.
 
In Tables~III and IV we give an example of a set of values for 
the partial wave amplitudes $R_{\sl LJ}$ and their phases 
$\delta_{\sl LJ}$ found by our $\chi^2$ fit to the data 
$\overline{p}p \rightarrow K^-K^+$, with $J_{\sl max}$ = 2.  
The normalization of $R_{\sl LJ}$ is such that, if the momentum {\sl p} 
in Eqs.(5) and (6) is expressed in GeV/{\sl c}, the cross section 
defined in Eq. (1) is in $\mu$b/srad. 
The corresponding $\chi^2$ values are those of Table~I for 
$J_{\sl  max}$ = 2. From these tables one notes that 
the two $J$ = 2 partial wave amplitudes give very significant 
contributions to the cross section at all momenta. 
Since we have not used any energy smoothing procedure in our analysis, 
and since the ambiguities do not permit a unique solution, 
the amplitudes necessarily carry substantial uncertainties.
However, independent of ambiguities, all fits require an important 
contribution from the $J$ = 2 partial wave amplitudes 
at all energies and show a need for adding $J$ = 3 amplitudes  
for $p_{\rm lab}$ above 0.9 GeV/{\sl c}. 
This statement can be made, while no theoretical bias  as to the energy 
behaviour of the amplitudes has been imposed on our fits. 

These results are consistent 
with the simple model analysis at  higher energies~\cite{TMK}.
This earlier work~\cite{TMK} used a diffractive
scattering model from a simple black or grey sphere which
could explain most of the features of the higher energy data 
($p_{\rm lab}$
above 1.5 GeV/{\sl c} for $\overline{p}p \rightarrow \pi^- \pi^+$, 
and above 1.0 GeV/{\sl c} for $\overline{p}p \rightarrow K^- K^+$).
In that model description of the data, the spin dependent forces
were assumed to act in the surface region only.
The idea was that since the central region was ``black'' no detailed 
information would escape from the central interaction region.
Only the more transparent surface region would provide the spin-forces 
giving the asymmetries of this annihilation  reaction.

We interpret the fact that very few partial waves are needed in the 
analysis to mean that the annihilation reaction 
$\overline{p}p \rightarrow K^-K^+$ 
is a very central process as
was previously~\cite{Klo}  found also for the reaction
$\overline{p}p \rightarrow \pi^-\pi^+$. 
Moreover, this analysis clearly shows that 
$\overline{p}p \rightarrow K^-K^+$ 
requires even less partial waves than 
$\overline{p}p \rightarrow \pi^-\pi^+$. 
It is possible that part of this effect is related to the
annihilation of an additional quark-antiquark pair and the 
creation of strange quarks needed to obtain a $K^-K^+$ 
final state. We note that this effect cannot be explained by 
the lower final momenta in the $K^-K^+$ system versus the 
$\pi^-\pi^+$ system. A check of the kinematics shows
that corresponding momenta in $K^-K^+$ and $\pi^- \pi^+$ 
systems at these energies are not very different. 
At the lowest antiproton momentum, 360 MeV/$c$, the kaon momentum 
is a factor 0.86 of the pion momentum. 
As the antiproton energy increases, this factor approaches 1. 
Therefore the small difference in the final meson momenta does 
not explain that $J_{\sl max}$ = 2 for $K^-K^+$ and $J_{\sl max}$ 
= 3 for $\pi^-\pi^+$. 
Finally, it is difficult to interpret annihilation ranges for these 
reactions since both reactions are strongly influenced by 
final state meson-meson interactions~\cite{Julich} 
as well as by effects of coupling to other annihilation channels.
Therefore, statements about annihilation ranges  will necessarily 
be strongly model dependent. 

In conclusion, the present experimental data and analyses of these 
data for both $\overline{p}p \rightarrow \pi^- \pi^+$ 
and $\overline{p}p \rightarrow K^- K^+$ reactions 
may help to build better models of these simple,
but very fundamental annihilation reactions. 
Two properties are essential for a successful model description 
of these two reactions. The annihilation model should be of short 
range and the model should allow for substantial $J$ = 2 partial wave 
contributions to $\overline{p}p \rightarrow K^- K^+$
while still allowing a significant $J$ = 3 partial wave for the 
$\overline{p}p \rightarrow \pi^- \pi^+$ reaction. 

The analysis described in this paper 
suggests the following future experiments necessary in
order to clarify further the understanding of 
the $\overline{p}p \rightarrow K^-K^+$ annihilation reaction: 

(i) Measurements of this annihilation reaction should be made at 
antiproton momenta closer to threshold. At very low energies 
we expect that even fewer partial wave amplitudes would contribute 
and the ambiguities of an analysis would be reduced. 

(ii) A further constraint on the analysis of this reaction would come
from data on the other spin observables for 
$\overline{p}p \rightarrow K^- K^+$
with longitudinal and/or transverse polarized beam and target,
for example data on the spin-correlations $A_{\sl ss}$ or $A_{\sl ls}$.
No spin-correlation observables have as yet been measured. We
propose the use of polarized antiprotons to obtain 
data on  the observables $A_{\sl ls}$ or $A_{\sl ss}$ for 
$\overline{p}p \rightarrow K^- K^+$.
This would allow to determine more accurately the various angular 
momentum amplitudes and contribute further to our understanding of this 
fundamental annihilation and hadronization process. 

\bigskip

We thank F. Bradamante for providing us with the LEAR data, and 
R. Timmermans for many useful discussions on the art of phase shift 
analyses. One of the authors (W.M.K.) is grateful to the University of 
South Carolina for its hospitality during his stay, when this 
work was initiated. This work was supported in part 
by NSF Grant Nos. PHYS-9310124 and PHYS-9504866.

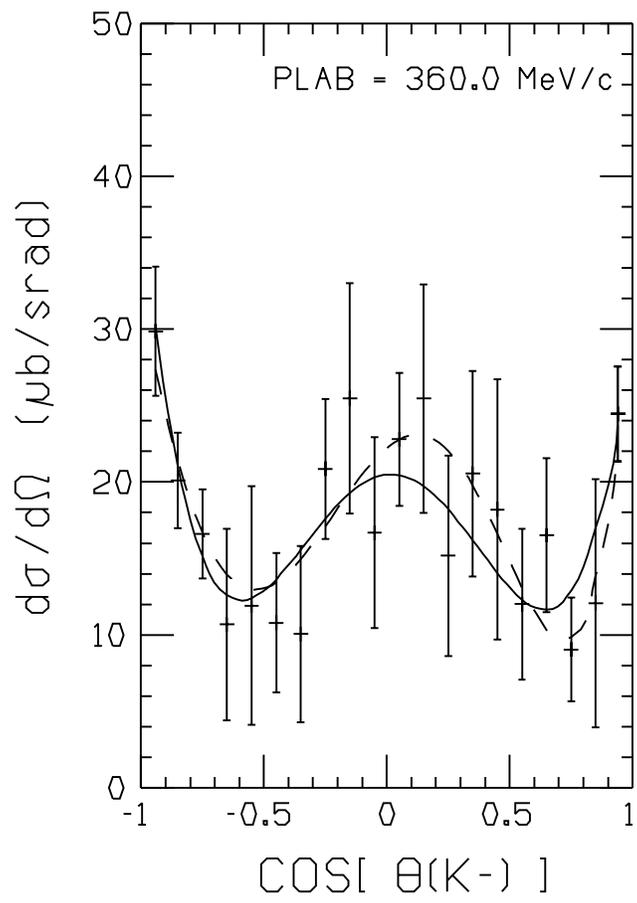
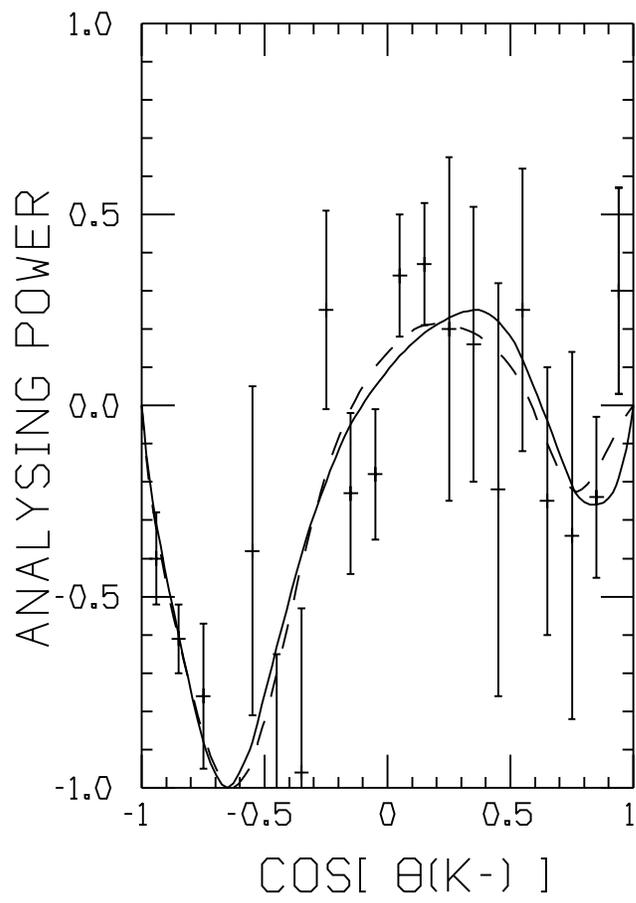
\begin{figure}
\caption[]{$\rm {\sl d}\sigma/{\sl d}\Omega$ and $A_{\sl on}$ at $p_{\rm 
lab}$ = 360 MeV/$c$ for $\overline{p}p \rightarrow K^-K^+$ .  
The solid curves give the fit for $J_{\sl max}$ = 2 and the 
dashed curves are the fit for $J_{\sl max}$ = 3.}
\end{figure}

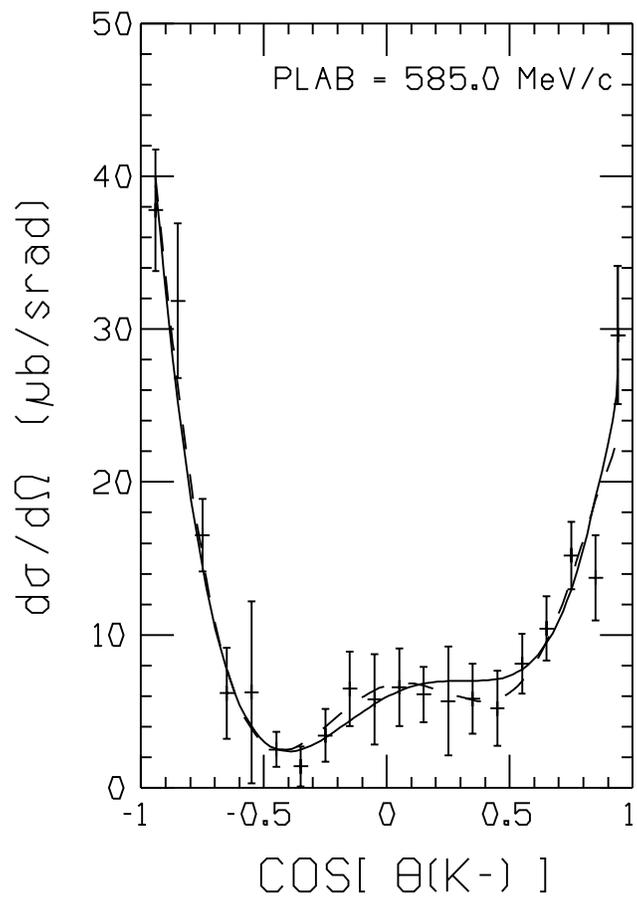
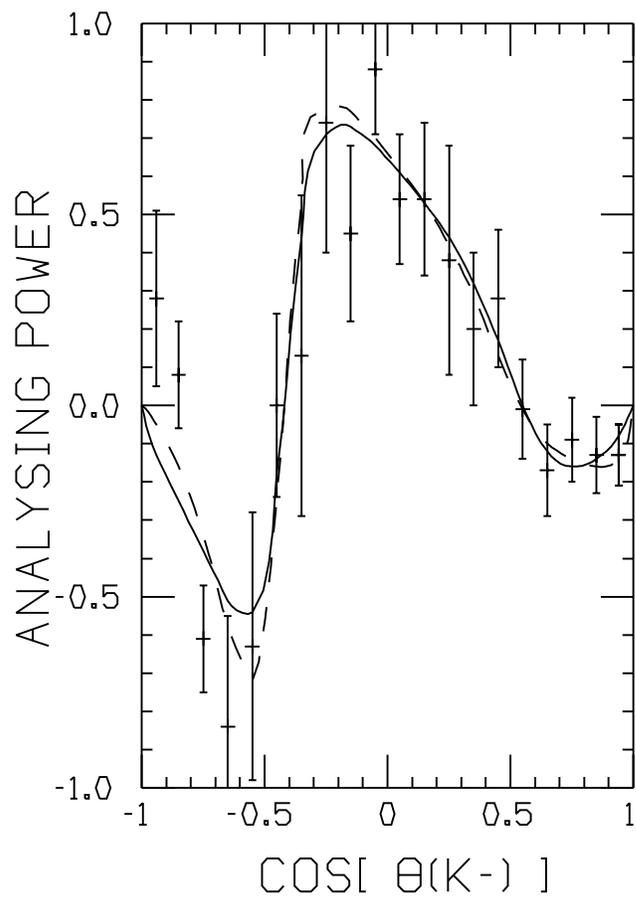
\begin{figure}
\caption[]{$\rm {\sl d}\sigma/{\sl d}\Omega$ and $A_{\sl on}$ at $p_{\rm 
lab}$ = 585 MeV/$c$ for $\overline{p}p \rightarrow K^-K^+$ .  
The solid curves give the fit for $J_{\sl max}$ = 2 and the 
dashed curves are the fit for $J_{\sl max}$ = 3.}
\end{figure}

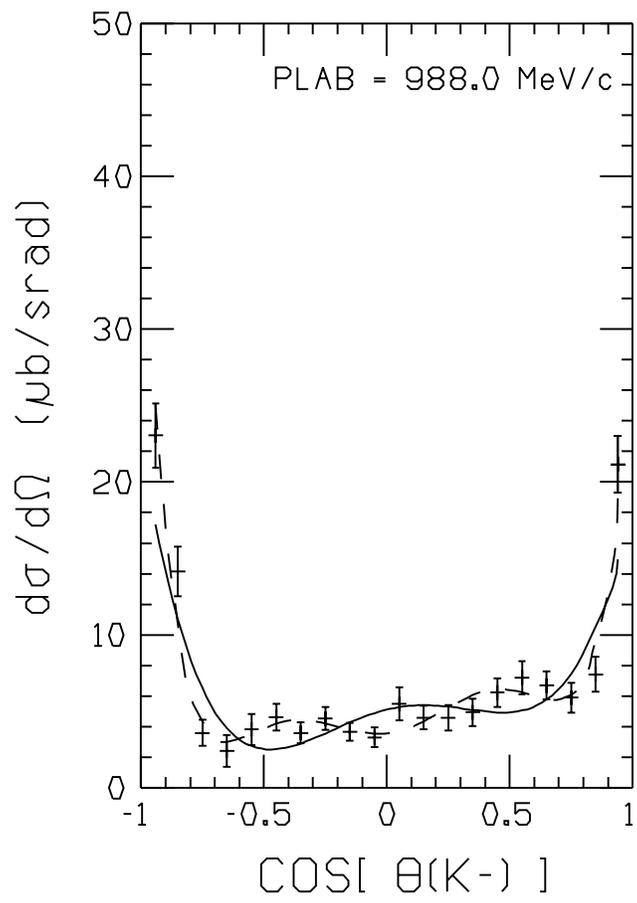
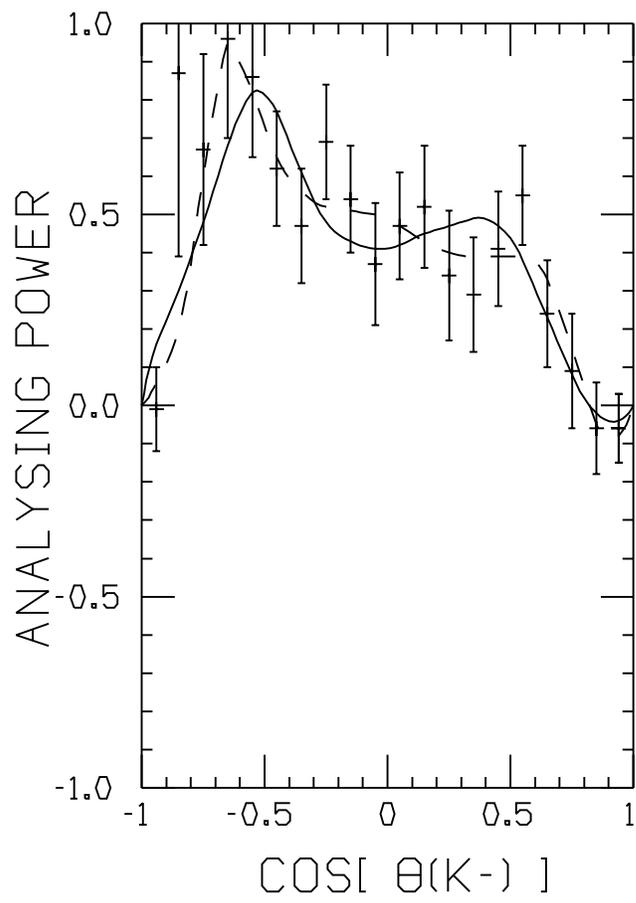
\begin{figure}
\caption[]{$\rm {\sl d}\sigma/{\sl d}\Omega$ and $A_{\sl on}$ at $p_{\rm 
lab}$ = 988 MeV/$c$ for $\overline{p}p \rightarrow K^-K^+$ .  
The solid curves give the fit for $J_{\sl max}$ = 2 and the 
dashed curves are the fit for $J_{\sl max}$ = 3.
For $p_{\rm lab}$ = 988 MeV/$c$ it is clear from $d\sigma/d\Omega$ 
that $J$=3 is necessary.}
\end{figure}

\squeezetable

\begin{table}
\caption{Examples of $\chi^2$ per degree of freedom for $K^- K^+$ 
at each energy.}
\begin{tabular}{ddddddddd} 
$p_{\rm lab}$ (MeV/$c$) & $\chi^2$($J_{\sl max}$=1)& 
$\chi^2$($J_{\sl max}$=2)& $\chi^2$($J_{\sl max}$ = 3) &
$\chi^2$($J_{\sl max}$ = 4)\\
\tableline
360&  2.26&  0.93&  0.97&  0.95\\
404&  2.32&  1.23&  1.40&  1.44\\
467&  1.90&  1.00&  1.13&  0.92\\
497&  3.81&  1.40&  1.02&  1.03\\
523&  4.70&  1.03&  0.72&  0.80\\
585&  3.25&  0.95&  0.98&  0.79\\
679&  3.42&  1.44&  1.53&  1.52\\
783&  6.92&  2.45&  2.41&  2.22\\
886&  5.08&  1.59&  1.37&  1.30\\
988&  4.73&  2.80&  1.06&  1.20\\
\end{tabular}
\end{table}

\begin{table}
\caption{Examples from Ref.\protect\cite{Klo} 
of $\chi^2$ per degree of freedom for $\pi^- 
\pi^+$ at each energy.}
\begin{tabular}{ddddddddd} 
$p_{\rm lab}$ (MeV/$c$) & $\chi^2$($J_{\sl max}$=2)& 
$\chi^2$($J_{\sl
max}$ = 3) &
$\chi^2$($J_{\sl max}$ = 4)\\
\tableline
360&  1.96&  1.77&  1.74\\
404&  1.38&  1.12&  1.12\\
467&  1.98&  1.31&  1.18\\
497&  3.04&  1.50&  1.45\\
523&  2.63&  1.45&  1.43\\
585&  1.96&  1.51&  1.57\\
679&  2.17&  1.50&  1.53\\
783&  2.50&  1.49&  1.47\\
886&  3.21&  1.23&  1.13\\
988&  4.39&  1.85&  1.55\\
\end{tabular}
\end{table}

\begin{table}
\caption{Energy dependence of parameters of $R_{\sl LJ}$ and
$\delta_{\sl LJ}$ for $J_{\sl max}$ = 2 for $K^- K^+$.}
\begin{tabular}{ddddddddd} 
$p_{\rm lab}$ (MeV/$c$) & $R_{10}$ & $\delta_{10}$ &
$R_{01}$ & $\delta_{01}$ &
$R_{21}$ & $\delta_{21}$
\\
\tableline
360&  0.94&  0 & 0.26&   35& 0.30& 75\\
404&  0.94&  0 & 0.22&   50& 0.24& 40\\
467&  1.38&  0 & 0.20&    5& 0.30& 70\\
497&  0.94&  0 & 0.22&   30& 0.26& 50\\
523&  1.08&  0 & 0.30&   20& 0.40& 60\\
585&  1.14&  0 & 0.20&  -50& 0.44& 50\\
679&  1.26&  0 & 0.56&  -85& 0.40& 35\\
783&  1.38&  0 & 0.58&  -95& 0.36& 35\\
886&  1.70&  0 & 0.66&  -95& 0.20& 45\\
988&  1.62&  0 & 0.36& -100& 0.18& 10\\
\end{tabular}
\end{table}

\begin{table}
\caption{Energy dependence of parameters $R_{\sl LJ}$ and
$\delta_{\sl LJ}$ for $J_{\sl max}$ = 2 for $K^- K^+$.}
\begin{tabular}{ddddddddd} 
$p_{\rm lab}$ (MeV/$c$) & $R_{12}$ & $\delta_{12}$ &
$R_{32}$ & $\delta_{32}$ 
\\
\tableline
360&  0.14&   75& 0.38& -90\\
404&  0.14&   95& 0.42& -85\\
467&  0.22&  125& 0.38& -65\\
497&  0.30&  105& 0.56& -55\\
523&  0.28&  130& 0.42& -50\\
585&  0.38&  125& 0.40& -60\\
679&  0.58&  130& 0.38& -55\\
783&  0.50&  130& 0.46& -50\\
886&  0.44&  110& 0.46& -40\\
988&  0.44&   95& 0.48& -65\\
\end{tabular}
\end{table}

\end{document}